\documentclass[preprints,article,accept,moreauthors,dvi2pdf]{Definitions/mdpi}

\firstpage{1} 
\pubvolume{xx}
\issuenum{1}
\articlenumber{5}
\pubyear{2020}
\copyrightyear{2020}
\history{Received: date; Accepted: date; Published: date}

\Title{NeuXus: A Biosignal Processing and Classification Pipeline for Real-Time Brain-Computer Interaction}

\Author{Athanasios Vourvopoulos $^{1,*,\dagger,\ddagger}$\orcidA{0000-0001-9676-8599}, Simon Legeay $^{2,\ddagger}$ and Patricia Figueiredo $^{1,\dagger}$}

\AuthorNames{Athanasios Vourvopoulos, Simon Legeay and Patricia Figueiredo}

\address{%
$^{1}$ \quad Institute for Systems and Robotics - Lisboa,
Instituto Superior Técnico, Universidade de Lisboa; \{athanasios.vourvopoulos, patricia.figueiredo\}@tecnico.ulisboa.pt\\
$^{2}$ \quad CentraleSupélec; simon.legeay@supelec.fr}

\corres{Correspondence: athanasios.vourvopoulos@tecnico.ulisboa.pt;}

\firstnote{Current address: Av. Rovisco Pais, 1, 1049-001 Lisboa, Portugal } 
\secondnote{These authors contributed equally to this work.}

\abstract{In the last few years, Brain-Computer Interfaces (BCIs) have progressed as an emerging research area in the fields of human-computer interaction and interactive systems. This is primarily due to the introduction of low-cost electroencephalographic (EEG) systems that render BCI technology accessible for non-medical research but also due to the advancements of signal processing and machine learning methods. Consequently, BCIs could provide a wide new range of possibilities in the way users interact with a computer system (e.g., neuroadaptive interfaces). However, major challenges must still be addressed for BCI systems to mature into an established communication medium for effective human–computer interaction. One of the major challenges involves the easy integration of real-time processing pipelines with portable EEG systems for an "out of the lab" use. To date, despite the amount of options current open-source tools provide, most toolboxes focus mainly in extending the processing and classification methods but lack on the ability to provide an easy-to-design yet extensible architecture for ubiquitous use. Here, we present NeuXus, a modular toolbox in Python for real-time biosignal processing and pipeline design. NeuXus is open-source and platform independent, providing high-level implementation of processing pipelines for easy BCI design and deployment.}


\begin{document}


\section{Introduction}
Brain-computer Interfaces (BCIs) can be described as communication systems able to establish an alternative pathway between the user’s brain activity and a computer, providing an additional non-muscular channel for communication and control to the external world \cite{wolpaw2002brain}. This can be achieved either explicitly by allowing users to issue direct commands into devices without physical involvement of any kind (e.g., speller typing, wheelchair control), or implicitly by monitoring a user’s state (e.g., workload level, attention state) to proactively adapt a user-interface or a virtual environment.

To date, the use of non-invasive Electroencephalography (EEG) is the most common brain signal acquisition technique in BCIs. EEG utilizes electrodes placed over the heads scalp for capturing the combined electrical activity of "firing" neurons. These neurons have characteristic intrinsic electrical properties, producing electrical and magnetic fields which propagate through volume conduction over the scalp \cite{da_silva_eeg:_2010}. Through this evoked electrical potential, different EEG patterns of brain activity are activated depending of the mental task. This oscillatory brain activity is currently used for the interfacing of humans with computers in a EEG-based BCI system. Some of the benefits of using EEG in BCIs include the high temporal resolution (1ms), EEG systems they are relatively lightweight and portable, and finally the cost compared to other brain imaging technologies (e.g., Magnetoencephalography), allows for more flexible data collection in real-world environments. On the downside, EEG signals are easily contaminated with artifacts and noise of non-brain sources, and have low spatial resolution, making source localization hard to achieve and rendering their utilization cumbersome \cite{lotte_review_2007}.

A unidirectional brain-to-computer communication is elicited first by a stimulus (visual, auditory or somatosensory), which is generating endogenous or exogenous potentials. Endogenous potentials are considered those that their occurrence is not related to the physical attributes of a stimulus (e.g. frequency or intensity), but to a person’s reaction to it. For example, motor-imagery BCI (MI-BCI), involves the imagination of movement and is considered a BCI paradigm that evokes endogenous potentials since users are generating motor-related brain patterns unrelated to the stimulus attributes \cite{PfurtschellerNeuper2001}. In contrast, BCIs using Steady State Visual Evoked Potentials (SSVEP) are considered an exogenous potential stimulus since it is caused by external visual stimulation of flashing lights which occur at the primary visual cortex of the brain \cite{Muller_Putz_2005}. In addition, P300 BCIs are also receive external visual simulations using brain responses that are generated approximately 300 ms after stimulus onset (hence the name P300). Nonetheless, P300 are considered to be also endogenous potentials since after the stimulus presentation, there is a stimulus validation process (a cognitive function) by the brain \cite{Sellers2012BcisTU}.  Both SSVEP and P300 paradigms are mostly used for patients in the locked-in state for communicating with the outside world \cite{kubler_braincomputer_2009}, while MI-BCI is capturing activity over the motor and somatosensory cortices and is utilized for motor restoration in neurorehabilitation \cite{biasiucci_brain-actuated_2018, vourvopoulos.fnhum.2019}.

Undoubtedly, BCIs have not only been proven to be important tools in the medical domain as either assistive or restorative interfaces but also have introduced a unique form for the human–computer interaction (HCI) paradigm \cite{graimann2010brain}. In the last few years, BCIs have progressed as an emerging research area in the fields of HCI and interactive systems, primarily due to the introduction of low-cost EEG systems that render BCI technology accessible for out of the lab research moving closer to the integration with wearable technology \cite{vourvop2019eeglass}. Consequently, BCIs provide a wide new range of possibilities in the way users interact with a computer system (e.g., neuroadaptive interfaces). However, major challenges must still be tackled for BCI systems to mature into an established communication medium for effective human–computer interaction.

One of the major challenges involves the easy integration of real-time processing pipelines with portable EEG systems for an out-of-the-lab use. To date, despite the amount of options current open-source tools provide, most toolboxes focus mainly in extending the processing and classification methods but lack on the ability to provide an easy-to-design yet extensible architecture for ubiquitous use. Here, we present NeuXus (\textbf{neu}ral + ne\textbf{xus}, latin for connection, pronounced IPA: /'nek.sus/), a modular toolbox in Python for real-time biosignal processing and pipeline design.

 \section{Background}
In EEG-based BCIs, specialized algorithms are responsible for translating EEG signals into control commands to devices in order to provide feedback. To date, a plethora of software tools for EEG processing are available
for both basic and clinical/translational research. This is highlighted in a significant amount of bibliometric reviews of the state-of-the-art of BCI software tools that go back up to 20 years of BCI research \cite{Stegman2020} \cite{Brunner_2011, Brunner2013} \cite{Delorme2010}.

Current software for EEG processing is highly diverse and it involves tools ranging from general purpose processing of electrophysiological signals up to more specific for BCI or neurofeedback use. In many cases, the application level (general vs specific) of the EEG/BCI tool dictates also the programming environment that is utilized for implementing the processing. For instance, the use of MATLAB is more common in most of the general-purpose biosignal or EEG signal processing tools for more than 20 years. On the other hand, first with C++ and next with Python, BCI-specific tools have emerged for consumer products and emerging applications. 

In MATLAB, a comprehensive set of EEG tools have been developed as toolboxes but also as standalone versions. Specifically, EEGLAB (Swartz Center for Computational Neuroscience, UCSD) \cite{DELORME20049} is one of the most widely used tools for EEG analysis, incorporating independent component analysis (ICA), time/frequency analysis, artifact rejection, event-related statistics, and several visualization features. 
Moreover, EEGLAB supPorts a large amount of plugins which extend its functionality. Specifically, for real-time EEG processing, the BCILAB plugin is supported for the design, prototyping, testing, experimentation and evaluation of BCIs \cite{kothe2013bcilab}. Similarly, Fieldtrip is also a MATLAB toolbox for EEG, MEG and NIRS analysis. Like EEGLAB, it offers pre-processing and advanced analysis methods, such as time-frequency analysis, source reconstruction using dipoles, and non-parametric statistical testing \cite{oostenveld_fieldtrip:_2011}. Furthermore, for real-time data acquisition the Fieldtrip buffer module is used. The FieldTrip buffer is a network transparent TCP server that allows the acquisition client to stream data to it per sample or in small blocks, while at the same time previous data can be analyzed. One of the oldest and still active EEG processing toolbox for MATLAB is Brainstorm (University of Southern California, Los Angeles, CA) \cite{Brainstorm2011}. Brainstorm is incorporating all the necessary processing tools for continuous or event-related data for MEG, EEG, fNIRS, ECoG, depth electrodes and multi-unit electrophysiology. For realtime data acquisition, Brainstorm is also using the FieldTrip buffer. Despite most of MATLAB toolboxes are available for free, MATLAB requires a paid license. Alternatively, the GNU Octave Project is a free alternative to MATLAB. Nonetheless, Octave does not support the entirety of the tools available (such as EEGLAB), with many of the functions not properly working, including the graphical user interface and menus. Some of the standalone processing tools include, BrainVision Analyzer (Brain Products GmbH), BrainVoyager (Brain Innovation B.V.) and LORETA (KEY Institute for Brain-Min Research, Zurich) for estimating cortical connectivity \cite{pascual-marqui_standardized_2002}. 

For specific BCI applications, specialized algorithms and signal processing libraries have been implemented in C++ for increased speed of execution and efficiency. Some of these tools include:  BCI2000 \cite{schalk2004bci2000}, one of the earliest BCI tools, TOBI (Tools for  Brain-Computer  Interaction)\cite{muller2011tobi}, BioSig \cite{schlogl2008biosig}, and OpenVibe (Inria, Rennes, France) \cite{renard2010openvibe}. Notably, OpenVibe is the only one using visual programming through a graphical interface, rendering it one of the most user friendly BCI tools, especially for entry level users.

In the last few years, many open-source Python software for exploring, visualizing, and analyzing EEG have emerged. For example, MNE-Python provides libraries for supporting the analysis of  MEG, EEG, sEEG, ECoG, NIRS data \cite{MNE-python}, being the mostly used Python tool for post-hoc analysis. Moreover, Wyrm is a BCI toolbox suitable for running on-line BCI experiments as well as, off-line analysis of EEG data together with Mushu \cite{Mushu} for signal acquisition and Pyff, a BCI feedback and stimulus framework \cite{Pyff}. Other tools in Python include Gumpy \cite{tayeb_gumpy_2018} and BCIPy\cite{memmott2020bcipy}. In most of these tools, the data acquisition support is extended through the use of the Lab Streaming Layer\footnote{https://github.com/sccn/labstreaminglayer} (LSL) protocol and the Extensible Data Format\footnote{https://github.com/sccn/xdf} (XDF)(Swartz Center for Computational Neuroscience, UCSD).

\section{Architecture and Functions}

NeuXus is a modular software written in Python 3\footnote{https://www.python.org/} and it requires version 3.7 (or greater) in order to be compatible with current dependencies and third-party libraries.
NeuXus processes are represented as Nodes in a graph structure, connected by edges called Ports. NeuXus Nodes can pass data to one another, interact with the operating system or a network stack.

\subsection{Nodes and Ports}
\begin{figure}[h]
\centering
\includegraphics[width=\textwidth]{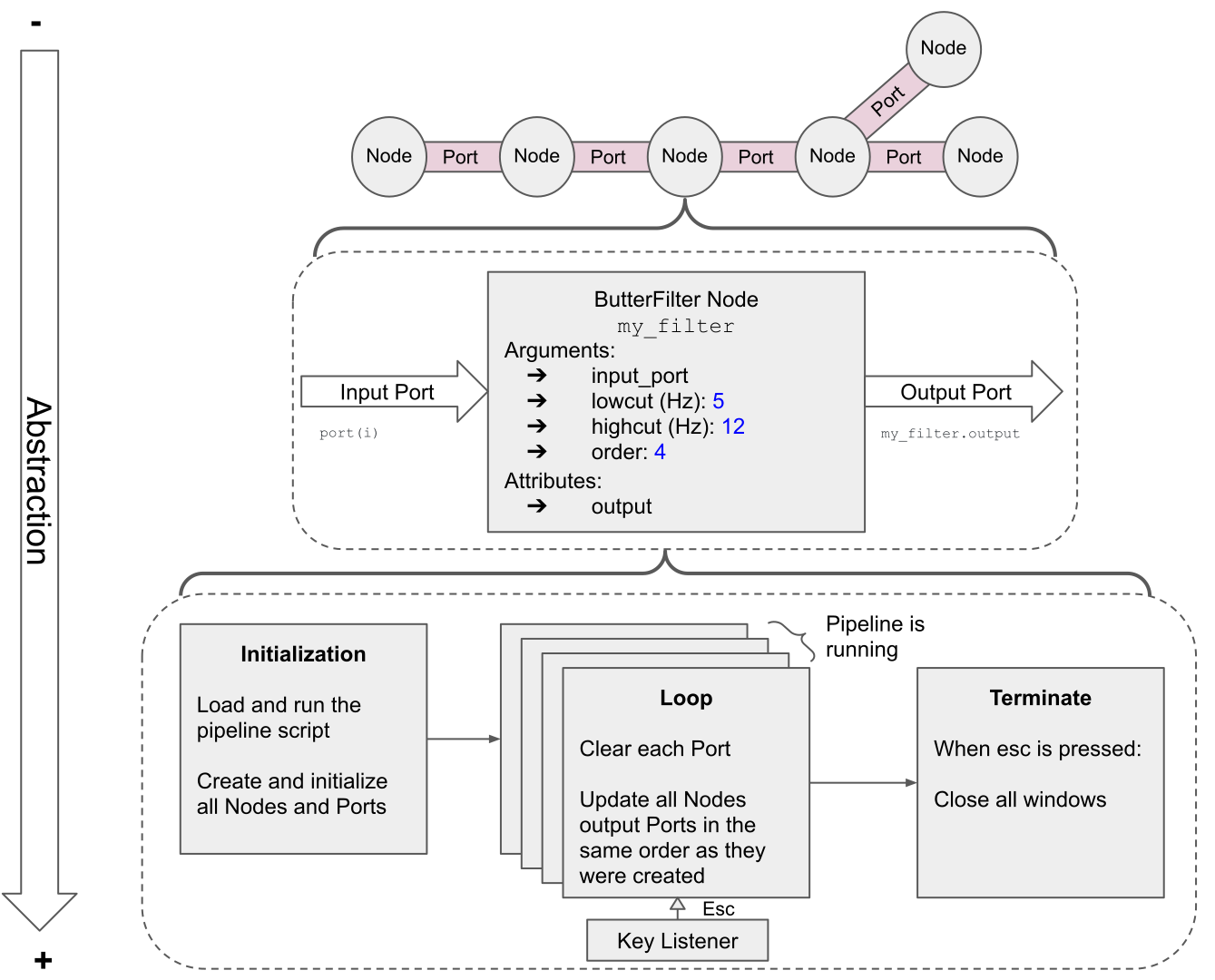}
\caption{NeuXus Nodes and Ports architecture in three abstracted levels}
\label{fig:neuxus arch}
\end{figure}

A Node represents a single function running sequentially in a NeuXus pipeline and managed by the pipeline.py script. Each Node has two ports, an input and an output (Figure \ref{fig:neuxus arch}). Every Node is instantiated with unique name and a set of parameters which are passed to the constructor of the Node during the initialization of the pipeline. 
During the initialization, Nodes are created, then each Node checks whether the input Port data type is coherent and creates the output port.
Next, it initializes the required attributes that are useful for the Node functionality. Finally, it creates a Log of each instance for debugging purposes.
After initialization, NeuXus runs the pipeline in a loop. On every loop, input Ports receive several chunk of data as a DataFrame\footnote{https://pandas.pydata.org/pandas-docs/stable/reference/api/pandas.DataFrame.html}, which is a two-dimensional data structure. Next, each DataFrame is processed based on the Node functions and finally it is forwarded to the output Port for the next node. All Nodes update their input and output Port(s) in the same order as they were instantiated in the pipeline.

\subsection{Functions}
Each Node represents a single function in a NeuXus processing pipeline. These processes are grouped under different modules based on their functionality (table \ref{tab:table1}). Current available modules or core modules are described in the following sections.

\subsubsection{IO}
The IO module is responsible for implementing input and output Nodes for real-time EEG or other biosignal data handling. 

The first supported Node is using LSL, a well established protocol for real-time data measurement and widely supported by the research community. Specifically, two LSL Nodes are available, one for streaming (LSLSend) and one for receiving data (LSLReceive) for both time-series and/or markers.

example:
\begin{lstlisting}[language=Python]
    #receive LSL stream
    lsl_signal_node = io.LslReceive('name', 'EEGAmp001', 
                                    data_type='signal', 
                                    sync='network')
    
    #send "lsl_signal" node output via LSL
    lsl_send_node = io.LslSend(input_port=lsl_signal_node.output,
                                name='my_lsl_signal',
                                type='signal')
    
\end{lstlisting}

Next, an RDA (Remote Data Access) Node is implemented (RdaReceive), supporting the data transferring protocol from Brain Products (Brain Products GmbH, Gilching, Germany) which allows to remotely access BrainVision Recorder for transferring data to other programs located on the local computer or in the network. The "RdaReceive" Node receives both signals and markers, while it can also be used as a bridge between BrainVision RecView and a NeuXus processing pipeline for the acquisition of real-time corrected EEG data from a simultaneous EEG-fMRI setup.

example:
\begin{lstlisting}[language=Python]
    #receive RDA data from a local computer with ip: 192.168.1.200
     rda_node = io.RdaReceive(rdaport=52136, 
                                offset=.125,
                                host='192.168.1.200')
\end{lstlisting}

Last but not least, data can be also be streamed via the UDP protocol by using the "UdpSend" node.

\subsubsection{Filter}
The filter module is responsible for the real-time signal filtering. Nodes are implemented using signal processing functions from the SciPy\footnote{https://www.scipy.org/} ecosystem. Specifically, a butterworth filter ("ButterFilter") for bandpass filtering, specifying the lowest and highest frequency cut in Hz and the order of the filter. A notch filter ("NotchFilter"), or band-stop filter which rejects a narrow frequency band (usually from power-line noise at 50 or 60 Hz) and leaves the rest of the spectrum little changed. Finally, a down-sampling Node ("DownSample") for changing the sample rate of the signal.

example:
\begin{lstlisting}[language=Python]
    #create a bandpass filter between 1-40Hz
    bandpass_node = filter.ButterFilter(input_port=port.output,
                                        lowcut=1,
                                        highcut=40,
                                        order=4)
\end{lstlisting}

\subsubsection{Processing}
The Processing module includes Nodes for further signal processing and transformations.
The "PsdWelch" Node is for Spectral Analysis using Welch’s Method, the "Fft" Node for Fast Fourier Transform and "HilbertTransform" Node for computing the analytic signal using the Hilbert transform.

example:
\begin{lstlisting}[language=Python]
    #Spectral Analysis of the bandpass_node
    psd_node = processing.PsdWelch(bandpass_node.output)
\end{lstlisting}

\subsubsection{Generate}
The Generate module hosts Node for artificial or synthetic data generation. The "Generator" Node is instantiated with one of three modes: 'random' for random time-series generation, 'oscillator' for sinusoidal signal generation, and 'simulation' for simulated EEG data. For every "Generator" node, the number of channels and the sampling frequency needs also to be specified.

\subsubsection{Read}
Read module hosts Nodes that load and replay data. The "Reader" Node is able to parse and replay data and markers in real-time. Specifically, it can replay EEGLAB set files (.set), General Data Format (.gdf), Extensible Data Format (.xdf), and Brain vision format (.vhdr).

\subsubsection{Select}
The Select module is responsible for data selection and spatial filtering. Specifically, the "ChannelSelector" Node allows to select outgoing signal data to a subset of incoming data based on a list of channels. Channels may be identified by their index (e.g. 1,5,8) or their name (e.g. Cz, Fp1, P3). The "SpatialFilter" Node maps M inputs to N outputs by multiplying the each input vector with a matrix of coefficients.
The "ReferenceChannel" Node subtracts the value of the reference channel from all other channels while the "CommonAverageReference" Node is subtracting from each sample the average value of the samples of all electrodes.

\subsubsection{Function}
The Function module hosts the "ApplyFunction" Node which can take as an input a python function and can pass a row or a numpy array (of shape number of input channels) for further processing inside a custom function.

example:
\begin{lstlisting}[language=Python]
    #function Node for subtracting 4 on every chunk of every channel
        def subfour(x):
            return x - 4
    
    applyfunction_node = ApplyFunction(port.output, subfour)
\end{lstlisting}

\subsubsection{Epoching}
The Epoching module hosts Nodes that are responsible for slicing (or epoching) the signal into chunks of a desired length following either a stimulation event or a specific time interval. Specifically, the "MarkerBasedEpoching" Node cuts a continuous signal each time a Marker appears without a specific name unlike the "StimulationBasedEpoching" Node which cuts the signal into chunks of a desired length of specific stimulation events (or stimulation codes). Last but not least, the "TimeBasedEpoching" Node cuts the signal in a specific time intervals and for a specific length. 

\subsubsection{Epoch function}
The Epoch Function module hosts Nodes that only process epoched data and not continuous. These Nodes can only be placed after an epoching node.
The "UnivariateStat" Node computes the 'mean', 'median', 'min', 'max', 'range', 'std', 'quantile', 'iqr' on each incoming epoched signal.
The "windowing" Node applies a windowing function between: 'blackman', 'hanning', 'hamming', 'triangular' on each incoming epoched signal.

\begin{table}[ht]
\caption {Core modules and nodes} \label{tab:table1} 
\begin{tabular}{ll}
\rowcolor[HTML]{F3F3F3} 
Module & Nodes \\
classify.py & Classify {[}LDA 'class' or 'probability'{]} \\
display.py & Plot, Graz, PlotSpectrum \\
epoch\_function.py & UnivariateStat, Windowing{[}'blackman', 'hanning', 'hamming', 'triang'{]} \\
epoching.py & TimeBasedEpoching, MarkerBasedEpoching, StimulationBasedEpoching \\
feature.py & FeatureAggregator \\
filter.py & ButterFilter, NotchFilter, DownSample \\
function.py & ApplyFunction \\
generate.py & Generator {[}'random', 'oscillator', 'simulation'{]} \\
io.py & LslSend, LslReceive, RdaReceive {[}'signal', 'marker'{]}, UdpSend \\
log.py & Hdf5, Mat \\
processing.py & PsdWelch, FFT, Hilbert \\
read.py & Reader {[}'.xdf', '.gdf', '.set', '.vhdr'{]}: \\
select.py & ChannelSelector, SpatialFilter, ReferenceChannel, CommonAverageReference \\
stimulator.py & Stimulator \\
store.py & ToCsv
\end{tabular}
\end{table}

\subsubsection{Display}
The Display module hosts Nodes responsible for signal visualization and feedback for stimulus markers. The "Plot" Node is visualizing signal streams and time series. The "PlotSpectrum" Node is for visualizing a spectrum signal in a matplotlib\footnote{https://matplotlib.org/} window. Finally, the "Graz" Node is used for rendering feedback related to the Graz BCI paradigm used for Motor Imagery BCI experiments\cite{pfurtscheller2001motor}. The "Graz" Node is using the Tk GUI toolkit\footnote{https://docs.python.org/3/library/tkinter.html} for rendering the stimulus as received through a stimulator node.

\subsubsection{Stimulator}
The Stimulator module is hosting Nodes responsible for generating stimulations based on experimental design parameters.
The "Stimulator" Node is generating a marker stream parsed from an external configuration file in xml format.

\subsubsection{Store and Log}
The Store module is for hosting Nodes responsible for data storage but not for large amounts of data like EEG datasets.
Currently, under the Store module, logging with comma separated values is supported through the "ToCsv" Node for logging low resolution time-series, events or markers.
For large datasets, the Log module hosts the "Hdf5" and "Mat" Nodes corresponding to the Hierarchical Data Format (HDF) and the Matlab matfile format respectively.

\subsubsection{Feature}
The Feature module is hosting Nodes responsible for creating feature vectors. The "FeatureAggregator" Node is taking each chunk of input and catenated into one feature vector together with class labels that can be used for classification.

\subsubsection{Classification}
The Classification module hosts Nodes responsible for model training and data classification. Currently the "Classify" Node is loading a trained model from joblib\footnote{https://pypi.org/project/joblib/} pipelining and assigns to the input vector either a 'class' or 'probability' value.

\subsection{Pipeline example}
NeuXus loads and executes a .py file (called the pipeline script) at the beginning of its execution, this file describes the pipeline. It initializes all Nodes and Ports between them.

\begin{lstlisting}[language=Python]
#This pipeline sends by LSL a simple DSP feedback calculated in real-time

from neuxus.Nodes import *

def square(x):
    # x is a matrix of a row
    return x ** 2

#Node 1: receive LSL stream
lsl_reception = io.LslReceive(
    prop='type',
    value='EEG',
    data_type='signal'
)
#Node 2: bandpass filtering between 8-12hz
butter_filter = filter.ButterFilter(
    input_port=lsl_reception.output,
    lowcut=8,
    highcut=12
)
#Node 3: square the signal from the square(x) function
apply_function = function.ApplyFunction(
    input_port=butter_filter.output,
    function=square
)
#Node 4: epoch 1 sec of data every 0.5sec
epoch = epoching.TimeBasedEpoching(
    input_port=apply_function.output,
    duration=1,
    interval=0.5
)
#Node 5: average signal
average = epoch_function.UnvariateStat(
    input_port=epoch.output,
    stat='mean'
)
#Node 6: send via LSL
lsl_send = io.LslSend(
    input_port=average.output,
    name='mySignalEpoched'
)
\end{lstlisting}

\subsection{Extending NeuXus}
\begin{figure}[h]
\centering
\includegraphics[width=\textwidth]{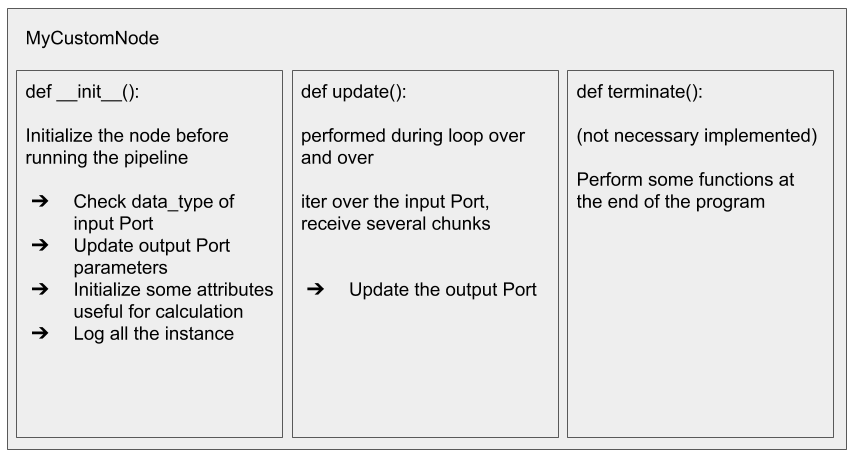}
\caption{Custom Node architecture}
\label{fig:Node arch}
\end{figure}

NeuXus is a modular software allowing developers and researcher to implement third-party or custom nodes. NeuXus loads first the pipeline script. It then repeats indefinitely a loop that clears all Ports and updates each node. The \textit{update} function of each Node is called at each iteration and get data from the input port. NeuXus pipeline is interrupted when 'Esc' is pressed.
A Node has then 3 methods \textit{init}, \textit{update} and \textit{terminate} (Figure \ref{fig:Node arch}).

A Node can have a port as input and/or output. Ports can share different type of data stored in Pandas DataFrame (table). A Port is an iteration object, it means that to get its values you need to iterate over it.

Type of Ports are:

\begin{itemize}
  \item 'signal': 1 iteration = 1 chunk of the continuous signal
  \item 'epoch': 1 iteration = 1 epoch
  \item 'marker': 1 iteration = 1 marker
  \item 'spectrum': 1 iteration = 1 spectrum computed at one timestamp
  \item 'vector': 1 iteration = 1 vector
\end{itemize}



\section{Discussion}
NeuXus is an open-source and platform independent biosignal processing tool that attempts to balance usability with functionality.
First, in terms of usability, NeuXus provides an abstract level of pipeline design through Nodes and Ports for real-time biosignal processing through its  core functions. Secondly, its functionality can be extended with custom code by creating new Nodes in Python. Last but not least, the processing pipelines of NeuXus can also exchange real-time data with external BCI tools through the supported networking protocols (e.g. LSL). Although interacting with NeuXus might not be as high-level as with visual programming tools (e.g. OpenVibe), nonetheless, and accounting for the trade-off between usability, functionality and extendability, NeuXus provides an easy-to-use interfacing with the Python language through the pipeline design.

In terms of performance, the speed of real-time processing is limited by the processing speed of each host that runs the pipeline. The processing frequency is self-regulated, hence, the software calculates all updates as quickly as possible and then waits for the next data entry. If the RAM is overloaded (because of other software or because there are too many Nodes in the pipeline), the loop takes longer to complete. This may involve a significant (and exponentially increasing) delay. If the stimuli is sent from outside the pipeline (e.g. via LSL), it is recommended to divide your pipeline, one with an independent stimulator, the other with the processing Nodes. This ensures that the pipeline does not slow down the sending of stimulation's via LSL. Further, by choosing Python, one loses execution speed compared to C++, nonetheless, each pipeline that can be designed with NeuXus can be considerably optimized on a case-by-case basis.


Last but not least, NeuXus is freely available through the Python Package Index (PyPI): \url{https://pypi.org/project/neuxus/} and the package can be installed using the pip package installer: \textit{pip install neuxus}. Finally, both the repository: \url{https://github.com/LaSEEB/NeuXus} and
documentation: \url{https://laseeb.github.io/NeuXus} are available online.



\vspace{6pt} 



\authorcontributions{Conceptualization, A.V.; software, S.L. and A.V; validation, S.L. and A.V; writing--original draft preparation, A.V.; writing--review and editing, P.F.; supervision, P.F.; All authors have read and agreed to the published version of the manuscript.}


\acknowledgments{The authors acknowledge the financial support of the Fundacao para a Ciencia e Tecnologia through FCT CEECIND/01073/2018, the LARSyS - FCT Project UIDB/50009/2020, the NeurAugVR project (PTDC/CCI-COM/31485/2017), and the MIG N2Treat Project (LISBOA-01-0145-FEDER-029675).}

\conflictsofinterest{The authors declare no conflict of interest.} 



\reftitle{References}


\externalbibliography{yes}
\bibliography{bibliography}




\end{document}